# Nucleation, growth, and dissolution of Ag nanostructures formed in nanotubular J-aggregates of amphiphilic cyanine dyes

Egon Steeg[a], Frank Polzer[a,c], Holm Kirmse[a], Yan Qiao[a,d], Jürgen P. Rabe[a,b], Stefan Kirstein[a]*

[a] Department of Physics, Humboldt-Universität zu Berlin, Newtonstr 15, 12489 Berlin, Germany

[b] IRIS Adlershof, Humboldt-Universität zu Berlin, Newtonstr 15 12489 Berlin

[c] present address: Materials Science & Engineering, University of Delaware, Newark, DE 19716, USA.

[d] present address: Centre for Protolife Research and Centre for Organized Matter Chemistry, School of Chemistry, University of Bristol, Bristol BS8 1TS, United Kingdom

Fax: +49-30-2093-7632; Tel: +49-30-2093-7714;
E-mail: kirstein@physik.hu-berlin.de

# Abstract

The synthesis of silver nanowires in solution phase is of great interest because of their applicability for fabrication of plasmonic devices. Silver nanowires with diameters of 6.5 nm and length exceeding microns are synthesized in aqueous solution by reduction of silver ions within the nanotubular J-aggregates of an amphiphilic cyanine dye. The time scale of the growth of the nanowires is of the order of hours and days which provides the unique possibility to investigate the nucleation, growth, and dissolution of the nanowires by direct imaging using transmission electron microscopy. It is found that the initial nucleation and formation of seeds of silver nanostructures occurs randomly at the outer surface of the aggregates or within the hollow tube. The growth of the seeds within the inner void of the tubular structures to nanowires indicates transport of silver atoms, ions, or clusters through the bilayer wall of the molecular aggregates. This permeability of the aggregates for silver can be utilized to dissolve the preformed silver wires by oxidative etching using $Cl^-$ ions from dissolved NaCl. Although the nanosystem presented here is a conceptual rather simple organic-inorganic hybrid, it exhibits growth and dissolution phenomena not expected for a macroscopic system. These mechanisms are of general importance for both, the growth and the usage of such metal nanowires, e.g. for plasmonic applications.

# Introduction

Metal nanostructures have become an interesting research topic[1] because of their applicability to the field of plasmonics.[2] The evanescent electric field of plasmon excitations within metallic nanostructures is used to enhance excitations of surrounding media and to localize optical excitations below the diffraction limit. Prominent examples are the usage of gold and silver nanostructures for field enhanced Raman spectroscopy, either by surface effects (SERS) or locally by tips of scanning probe microscopes (TERS). For particular applications the shape of the metallic nanostructures has to be controlled in order to tailor the geometry of the evanescent field, where high aspect ratio rod-like particles are favored geometries. In addition to lithographic methods, many synthetic routes have been developed to produce rod-like metal nanoparticles by solution processing. Two methods are widely used to control anisotropic growth that leads to rod-like or wire-like structures: Either the different growth speed of differently facetted surfaces is amplified by respective capping agents (polyol synthesis)[3], or templates are used to confine the growth. The latter method has been applied using hard templates such as zeolites[4] or soft templates such as the micelles of the surfactant CTAB.[5,6] The soft templating method allows solution synthesis of regular silver nanorods with diameters typically on the order of several ten nanometers and lengths over a hundred microns.[7]

A peculiar method to produce silver nanowires with diameters of 6.5 ± 1 nm was recently discovered using tubular J-aggregates as soft templates[8]. The aggregates self-assemble in aqueous solution from an amphiphilic cyanine dye (C8S3, 3,3′-bis(2-sulfopropyl)-5,5′,6,6′-tetrachloro-1,1′-dioctylbenzimidacarbocyanine) and form very regular and uniform tubular structures with an outer diameter of 13 ± 1 nm, and an inner diameter of 6.5 ± 1 nm while their lengths exceed microns.[9,10,11] The tube wall consists of a dye molecule bilayer, similar to a lipid bilayer, with a thickness of roughly 3 nm. These aggregates exhibit a strong J-aggregate character with red-shifted absorption and emission spectra. The spectra consist of several excitonic transitions, are very sensitive to structural changes, and can be used as a fingerprint to probe the integrity of the aggregate structure. Silver nanowires were produced by addition of $AgNO_3$ to solutions of the tubular J-aggregates and illumination by blue light for a short period of time [8].

Although various details of the wire structure and wire growth have been reported earlier, basic questions about the nucleation, growth, and material transport remained open. In general, the formation and growth of silver crystals in solution is described by a 3-step process:[12,13] First, nuclei are formed which consist of only a few atoms and are unstable. Second, these nuclei continue to grow into stable seeds which then, third, will grow further into crystallites. The primary step, the formation of nuclei, is of utmost importance for the morphology of the resulting crystallites.[14] Here, the time scale of the growth process is of the order of hours to days which provides the unique possibility to investigate the nucleation, growth, and dissolution of the nanowires by direct imaging using transmission electron microscopy (TEM) and cryo-TEM imaging.

# Results and discussion

## Nucleation and growth

Nucleation and growth of the nanostructures cannot be observed directly *in-situ*, but *ex-situ* the appearance of seeds and the growth of the crystallites can be followed. Parts of the solution were extracted at certain time steps after the illumination with light and investigated by TEM. The earliest time step after which a sample could be imaged was approx. 3 minutes, i.e. the time it took to prepare the TEM grids. At this time small seeds and crystallites have already been formed. Inspection of cryo-TEM images at this early stage reveals mostly bare aggregates with only few small silver crystallites (Fig. 1). Some of the nanocrystals are attached to the aggregates and in rare cases small pieces of nanowires grown inside the aggregates are found (marked in fig 1 by circle).

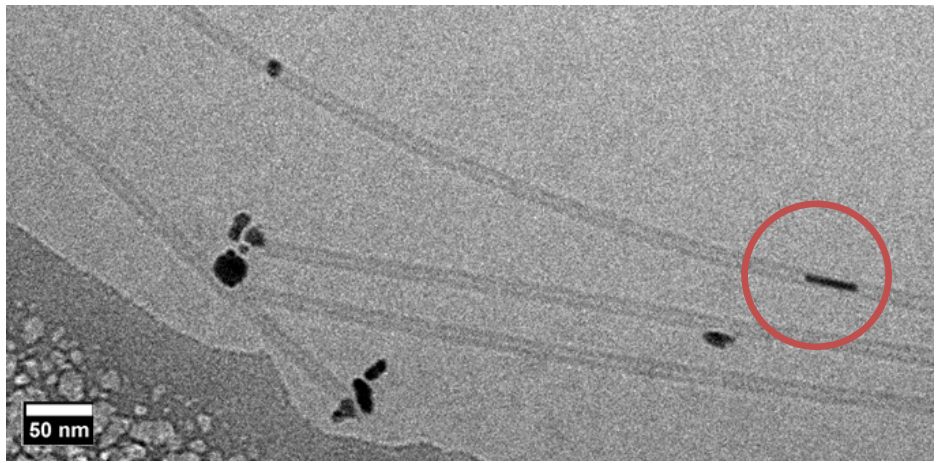

**Fig. 1:** Representative cryo-TEM image of tubular J-aggregates of C8S3 and few silver nanoparticles. A small piece of silver wire growing in the aggregate is found (marked by circle). The image was taken from a sample prepared for cryo-TEM immediately after addition of AgNO$_3$.

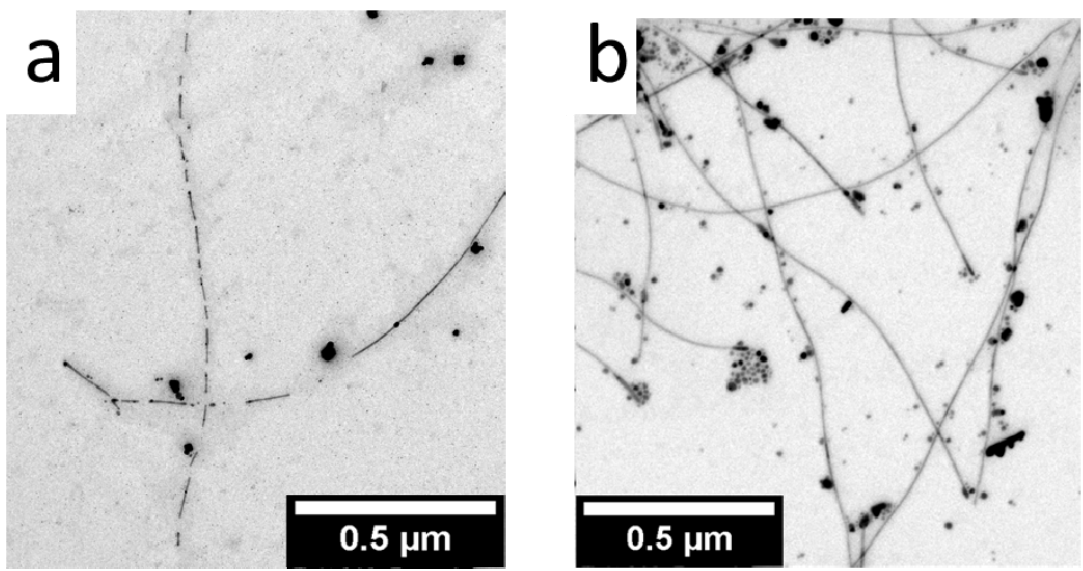

**Figure 2**: TEM images representing typical situations taken after 2 h (a) and 24 h (b) of growth time.

The main phase of growth from seeds to crystallites starts after few minutes and extends over several hours. Typical growth phases are depicted in Figure 2. While at short times only very short pieces of nanowires are found sporadically (Fig. 1), extended fragments of wires are visible after 2 hours (Fig. 2a). The length of the fragments has a broad distribution in the range of 20 – 500 nm and they are always oriented along traces. It is obvious that the fragments are confined within the tubular aggregate structures, but the molecular aggregates are not visible in TEM due to the low contrast. Inspection of manifold images supports the suggestion that every fragment of wire was growing from a single seed. Obviously, seeds are created along the aggregate at large distances (more than 50 nm) and then grow until the pieces connect to each other to form a continuous wire. Such situation is depicted in Figure 2b where wires with homogeneous thickness are seen with length exceeding microns. No short fragments interrupted by empty space are found.

Initially, i.e. immediately after the illumination of the solution with light, seeds must have developed from nuclei in the interior and / or at the outer side of the aggregates. The growth of the seeds into crystallites appears on the expense of silver ions, neutral silver atoms, or small silver nuclei. These atomically small reagents must be able to pass through the membrane-like wall of the dye tube. Otherwise, one cannot explain why the silver nanowires are growing inside the tubes. The number of ions present within the inner space after formation of the dye tube is not sufficient to form these structures and transport through the openings from the end pieces would be blocked by any piece of wire. Hence, material transport must be enabled by the permeability of the aggregate wall for silver ions, atoms, and clusters. The growth of crystals in the interior of the tubes, which leads to growth of the nanowires, obviously has to consume the silver not only from the interior volume, but also from a region outside the tube and hence lowers the probability for seed formation and growth of nanocrystallites at the outer surface of the aggregates.

If the growth time continues over days in the absence of light, the preformed nanowires grow also in diameter and expand across the size of the aggregates. This "over-growth" is due to the excess of silver ions still present in solution. The resulting crystals are wire-like, but are inhomogeneous and non-uniform in diameter (Fig. S4).

## Dissolution by oxidative etching

In order to prevent the over-growth, excess $Ag^+$ ions have to be removed from the solution. Therefore, NaCl was added in order to precipitate silver as AgCl by the reaction

$$Ag^+(aq) + Cl^-(aq) + Na^+(aq) \rightarrow AgCl(s) + Na^+(aq).$$

The molar amount of NaCl must be of the same order of magnitude or higher than the original amount of AgNO$_3$ to remove efficiently most of the unbound $Ag^+$. The interaction of the NaCl with the dye aggregates should be negligible, since NaCl does not change the pH of the solution and the dyes themselves are dissolved from a powder with Na as counter ions. Indeed it was observed that the removal of $Ag^+$ freezes the growth of silver wires at least for some time. Since the loss of absorbance of the J-bands is a good indicator for oxidation of the template and the growth of the silver nanostructures it can be used to observe the growth stop.

The growth of the wires is accompanied by changes in the optical absorption spectra (Fig. 3a). It is known from previous experiments[8,15] that within the first few minutes after illumination the optical absorption spectrum of the aggregates qualitatively changes and quantitatively decreases. The change in shape is due to the faster reduction of the highest absorption band at 590 nm, which is ascribed to absorption located at the outer dye layer of the aggregate, while the absorption band at 600 nm belongs to absorption located at the inner dye layer of the tubular aggregates. These changes can be understood by oxidation of the dye molecules[16], where dye molecules in the outer wall of the aggregates are oxidized preferentially[15]. The oxidation of the dyes only modifies the conjugated pi-electron system but does not destroy the molecules completely[17,16], which leaves the morphology of the aggregates unaffected.

In Figure 3 the absorption spectra are compared for a sample without any addition of NaCl (Fig. 3a) and for a sample, where NaCl was added 2 hours after initialization of the crystal growth by illumination. Therefore, the aggregate spectra at the time step of 2 hours are identical in Figure 3a and 3b. The NaCl was added in the same molar concentration as the $AgNO_3$. An increased absoption for wavelengths less than 300 nm is observed immediately after addition of NaCl, which is attributed to the absorbance of AgCl.

In the case where no NaCl is added the aggregates are almost completely oxidized after 6 days. In contrast, the sample with NaCl added still shows significant J-band absorption even after 6 days, although the shape of the spectrum has changed. An additional peak occurs at about 605 nm and at 575 nm a broad peak is seen that may contain several transitions. It is important to note that the total integrated absorbance remains constant for the whole time. The preservation of the J-Band over longer times indicates a prevention of further oxidation of the J-aggregate template and hence suppression of silver nanocrystal growth.

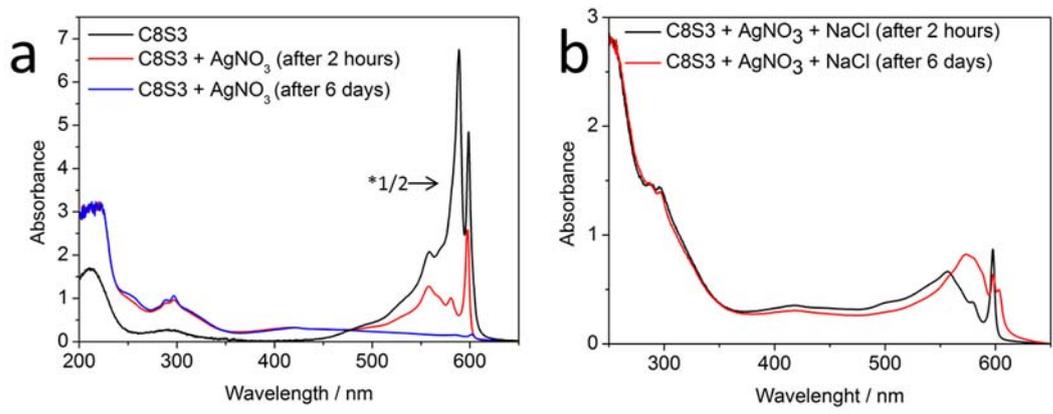

**Figure 3:** Absorption spectra of a) pure C8S3 solution (scaled by a factor 1/2), a solution measured 2 hours and 6 days after addition of AgNO3 b) same solution with $AgNO_3$ and same time steps as in a), but with the addition of NaCl after 2 hours.

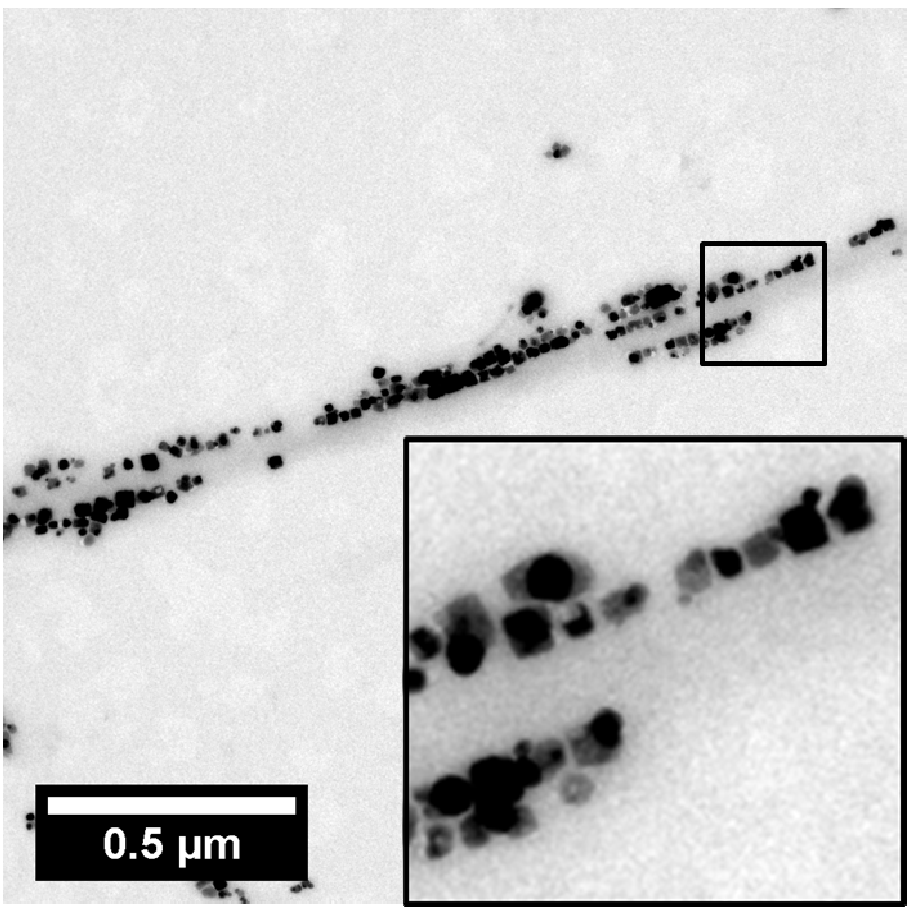

**Figure 4**: TEM image of a sample prepared from a C8S3/silver solution, where NaCl was added 2 hours after start of silver nanoparticle growth; the sample was stored for 2 days after addition of NaCl before imaging. The insert shows magnified view of marked area.

Inspection of samples by TEM reveals that shortly after addition of NaCl the structure of the silver wires is not changed. The wires show the typical features as seen in Figure 2a, namely, pieces of wires with length of several tens of nanometers interrupted by empty parts of the aggregates. A surprising result was found when the samples were imaged two days after the addition of the NaCl. No silver wires were found anymore; instead, aggregates were found covered by cubic crystals which can be identified as AgCl crystals (Fig. 4, and Fig. S3 for chemical analysis). One has to conclude that most or even all of the nanowires have been dissolved and the silver is converted into AgCl crystals. Such dissolution by Cl⁻ ions was reported recently for other silver nanocrystals and is explained by oxidative etching due to the presence of Cl⁻ as a ligand and $O_2$ in the solution [13]. This etching process was found to be more effective for nanostructures containing multiple twin boundaries and was used to synthesize single crystalline structures in high yield [18]. Therefore one may conclude that the crystalline silver nanowires presented here contain twin boundaries, possibly in high density. Indeed, studies of the crystal structure, which are beyond the scope of this paper, support this assumption.

## Conclusions

Silver nanowires grown in tubular J-aggregates provide a unique system of quasi one-dimensional structures, where the whole growth process can be observed by direct imaging techniques. It was found that the growth within such an amphiphilic template is controlled by transport of silver atoms, ions, or cluster through the bilayer wall of the tube. This material transport through the confining wall of the tube enables dissolution of the nanowires by oxidative etching. Although the system presented here seemed to be a simple two-component system, i.e. neutral silver and the templating aggregate, it exhibits growth and dilution phenomena which are caused by the size in the nanometer range and not expected for a macroscopic system.

## Experimental section

### Preparation of J-Aggregates

The cyanine dye 3,3′-bis(2-sulfopropyl)-5,5′,6,6′-tetrachloro-1,1′-dioctylbenzimidacarbo-cyanine (C8S3) was obtained from FEW Chemicals as a sodium salt with molecular weight 902.8*g mol$^{-1}$ and used as received. The double-walled nanotubular J-aggregates were prepared via the so called *alcoholic route*[19] in a water/methanol mixture (100:13 by volume) resulting in a final dye concentration of $c_{C8S3} = 3.36*10^{-4}$ M. Further experiments had to be performed within 24h of solution preparation.

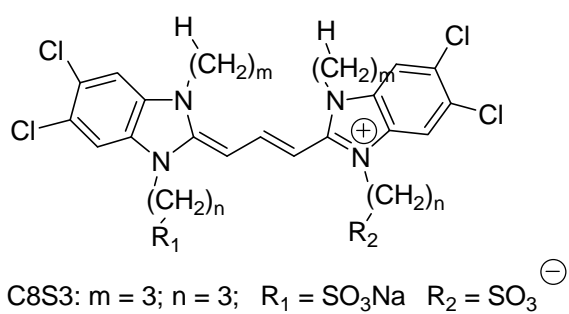

C8S3: m = 3; n = 3; R$_1$ = SO$_3$Na  R$_2$ = SO$_3^{\ominus}$

**Scheme 1**: Chemical structure of amphiphilic cyanine dye 3,3′-bis(2-sulfopropyl)-5,5′,6,6′-tetrachloro-1,1′-dioctylbenzimidacarbocyanine (C8S3)

### Preparation of silver nanowires and addition of sodium chloride

Silver nitrate at a concentration of 100 mM disolved in water was added to a freshly prepared J-aggregate solution (11:400 by volume) in a glass vial (Fisher Scientific) and then illuminated by 420 nm light for 1 minute employing a Xenon lamp of a Jasco FP-6500 fluorescence spectrometer. During growth of the silver nanostructures the samples were stored in the dark. Sodium chloride at a concentration of 100 mM was added to a silver nitrate/C8S3 solution with the same volume concentration as the silver nitrate (11:411 by volume).

## Absorption spectroscopy

Absorption spectra were taken with a double-beam UV-Vis spectrometer Shimadzu UV-2101PC. Quartz cells (Hellma GmbH) with path lengths of 0.2 mm and 2 mm were used.

## Transmission electron microscopy (TEM)

Conventional TEM measurements and energy-dispersive X-ray spectroscopy (EDXS) were performed with a JEOL JEM2200FS microscope at a beam energy of 200 keV (field emission gun). Samples were prepared by absorbing a small (5μL) droplet of solution on a 400-mesh copper TEM grid (Plano GmbH). Prior to sample deposition, the grids were hydrophilized by storing them over a water bath for 24 hour and blotted with a filter paper after 10 minutes. Cryo-TEM was performed with a JEOL JEM2100 at a beam energy of 200 keV ($LaB_6$ cathode). Droplets of the sample (5μl) were applied to perforated (2 μm hole diameter) carbon film grid (Quantifoil R2/2 200 mesh). In general conventional TEM allows for easier and faster preparation of the samples, therefore it is favored to screen a large quantity of samples in order to obtain better statistics. Cryo-TEM preserves the structure of the J-Aggregates but preparation is more challenging. Both techniques were applied whenever it was appropriate.

# Acknowledgements

We thank Evi Poblenz, Humboldt-Universität, for assistance. Dr. F. Polzer gratefully acknowledges the Joint Lab for Structural Research Berlin within IRIS Adlershof for funding. We thank the Deutsche Forschungsgemeinschaft (collaborative research center CRC 951) for financial support.

# Nucleation, growth, and dissolution of Ag nanostructures formed in nanotubular J-aggregates of amphiphilic cyanine dyes

# Supporting Information

*Egon Steeg[a], Frank Polzer[a,c], Holm Kirmse[a], Yan Qiao[a,d], Jürgen P. Rabe[a,b], Stefan Kirstein[a]\**

[a] *Department of Physics, Humboldt-Universität zu Berlin, Newtonstr 15, 12489 Berlin, Germany*

[b] *IRIS Adlershof, Humboldt-Universität zu Berlin, Newtonstr 15 12489 Berlin*

[c] *present address: Materials Science & Engineering, University of Delaware, Newark, DE 19716, USA.*

[d] *present address: Centre for Protolife Research and Centre for Organized Matter Chemistry, School of Chemistry, University of Bristol, Bristol BS8 1TS, United Kingdom*

Fax: +49-30-2093-7632; Tel: +49-30-2093-7714;
E-mail: kirstein@physik.hu-berlin.de

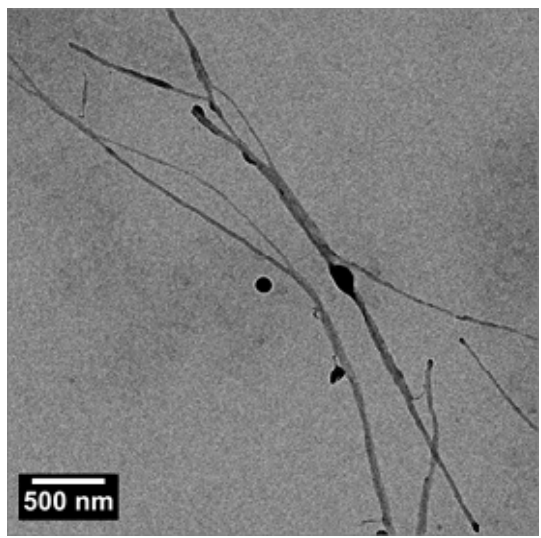

**Figure S1:** TEM image of a sample that has grown for 6 days. The silver structure has grown to in diameter across the template diameter, leading to irregular structures with inhomogeneous thickness.

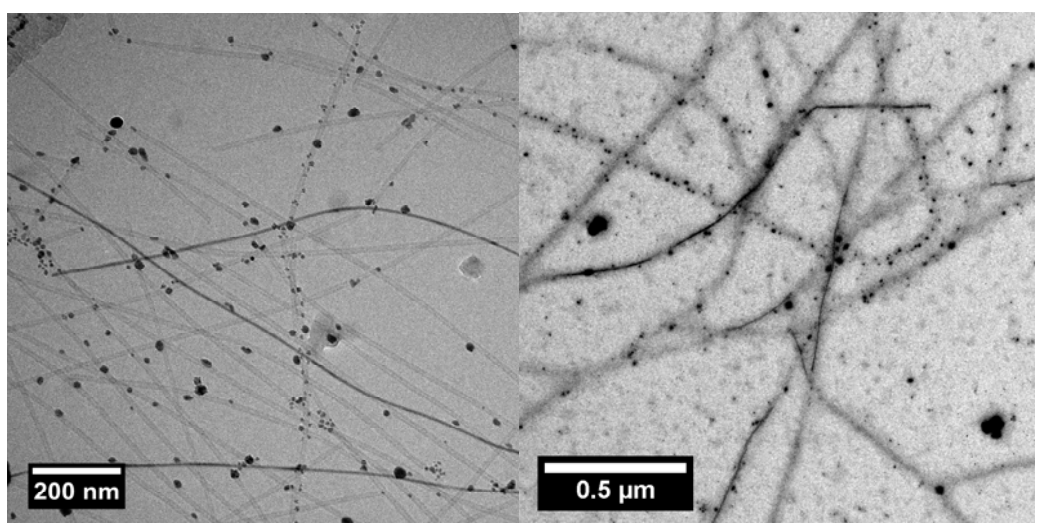

**Figure S2**: a) Cryo-TEM image of C8S3 aggregates after addition of AgNO3. The sample was prepared on a TEM grid 10 minutes after addition of the AgNO3. The coexistence of plane aggregates, aggregates decorated by Ag nanoparticles, and aggregates filled with Ag nanowires is visible. b) TEM image of the same solution as presented in a). The bare aggregates appear as blurred gray shaded traces.

Detailed Energy-dispersive X-ray spectroscopy (EDXS) investigations shown in **Figure S3** prove the composition of silver (red) and chlorine (green) for these structures.

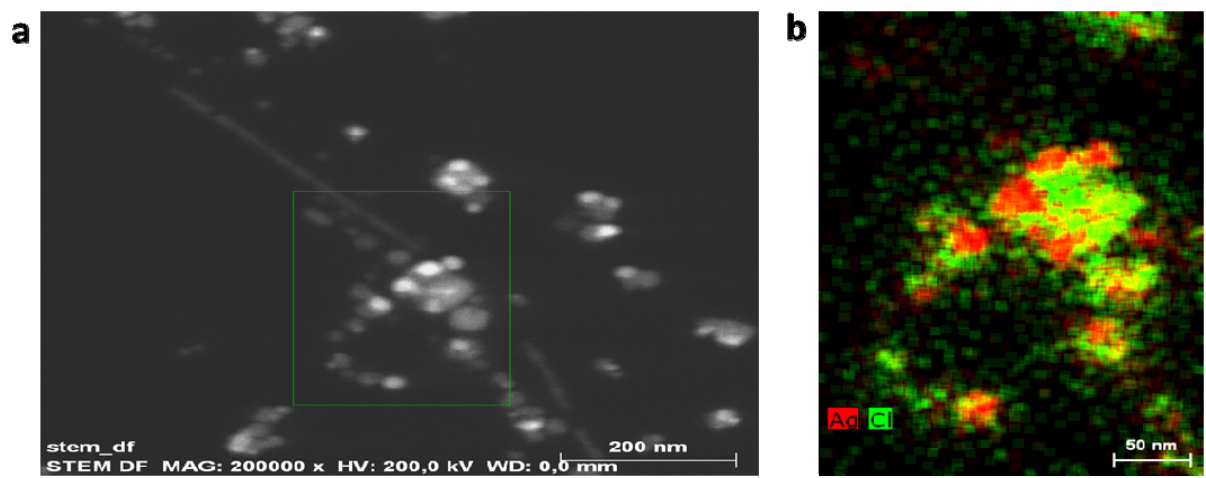

**Figure S3**: **a)** Scanning-TEM dark field image of a C8S3 solution with AgNO3 and addition of NaCl after 10 min **b)** Energy-dispersive X-ray spectroscopy of the area marked in a) with the contribution of silver in red and chlorine in green

**Figure S4a** shows sample 1 after a growth time of 2 days. Only cubic structures are visible. The particles are arranged along the surface of an elongated structure. Due to the very low contrast of the blurred trace and the comparison with the other TEM images it is expected that this is the J-aggregate. During the sample preparation for the TEM the solution is dried on a carbon film which leads to distorted aggregates. To make sure that the structure observed in the TEM is not an artefact due to the sample preparation Cryo-TEM has also performed on the solution. **Figure S4b** shows a Cryo-TEM image of sample 1 after 2 days. Here the J-Aggregate is clearly visible. The particles are decorating the template without disturbing its morphology.

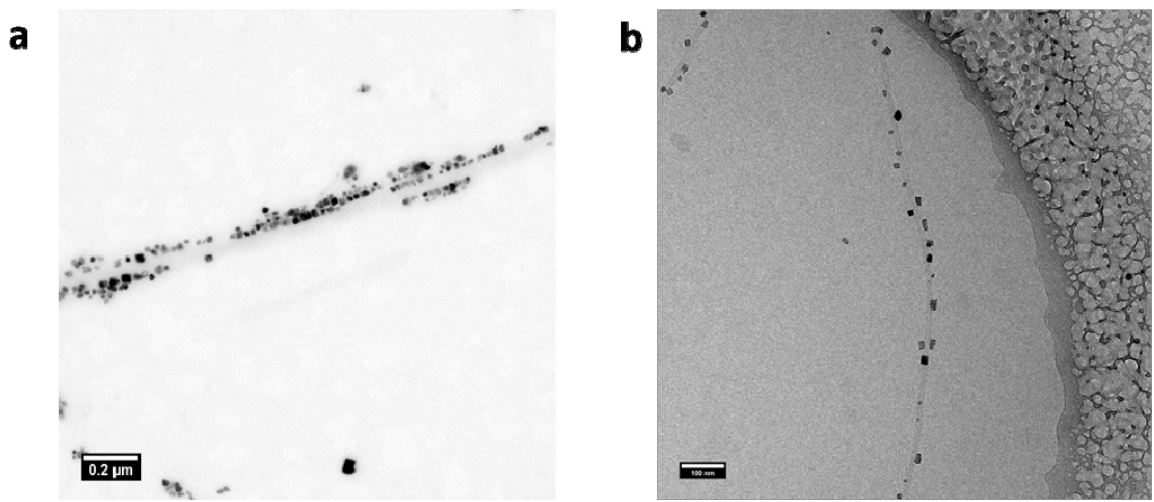

**Figure S4**: **a)** TEM image of a C8S3 solution with AgNO3 and addition of NaCl after 2 days **b)** Cryo-TEM image of the same sample as in a)